\documentclass[a4paper,10pt,aps,prd]{revtex4-1}
\usepackage[utf8]{inputenc}
\usepackage[title]{appendix}
\usepackage[colorlinks=true,allcolors=black]{hyperref}

\usepackage{amssymb}
\usepackage{mathtools}

\DeclareMathOperator{\STr}{STr}
\usepackage{breqn}
\usepackage{multirow}

\numberwithin{equation}{section}

\makeatletter
\newcommand{\hx}{\hat{\xi}}
\newcommand{\hs}{\hat{\sigma}}
\newcommand{\mS}{\mathcal{S}}
\newcommand{\tS}{\tilde{\mathcal{S}}}
\newcommand{\bl}{\tilde{\lambda}}

\newcommand{\bR}{\bar{R}}
\newcommand{\bg}{\bar{g}}
\newcommand{\bC}{\bar{C}}

\graphicspath{{plot/}}

\makeatletter
\let\cat@comma@active\@empty
\makeatother

\begin{document}

\author{Alfio Bonanno}
\email{alfio.bonanno@inaf.it}
\affiliation{INAF, Osservatorio Astrofisico di Catania, 
via Santa Sofia 78, 
I-95123 Catania, Italy;\\
INFN, Sezione di Catania, via Santa Sofia 64, 
I-95123,Catania, Italy.}

\author{Giovanni Oglialoro}
\email{giovanni.oglialoro@dfa.unict.it}
\affiliation{Dipartimento di Fisica e Astronomia “Ettore 
Majorana”, Universit\`a di Catania, 64, Via S. Sofia, I-
95123 Catania, Italy;\\
INAF, Osservatorio Astrofisico di Catania, 
via Santa Sofia 78, 
I-95123 Catania, Italy;\\
INFN, Sezione di Catania, Via Santa Sofia 
64, I-95123 Catania, Italy;\\
Centro Siciliano di Fisica Nucleare e 
Struttura della Materia, Catania, Italy.}

\author{Dario Zappal\`a}
\email{dario.zappala@ct.infn.it}
\affiliation{INFN, Sezione di Catania, Via Santa Sofia 
64, I-95123 Catania, Italy;\\
Centro Siciliano di Fisica Nucleare e 
Struttura della Materia, Catania, Italy.}

\begin{abstract}
\vskip 30pt
\centerline{ABSTRACT}
\vskip 10pt
Proper time functional flow equations have garnered significant 
attention 
in recent years, as they are particularly suitable in analyzing 
non-perturbative contexts.
By resorting to this flow, we investigate  the regulator 
and gauge dependence in quantum Einstein gravity within the 
asymptotic 
safety framework, considering various regularization schemes. 
Our findings indicate that some details of the regulator have
minor influence on the critical properties of the theory. In 
contrast, 
the selection between linear and exponential parametrizations 
appears to have 
a more substantial impact on the scaling behavior of the 
renormalized flow  near the non-Gaussian fixed point. 
\end{abstract} 


\title{Gauge and parametrization dependence of Quantum Einstein Gravity within 
the Proper Time flow}

\maketitle

\setcounter{page}{2}
\section{Introduction}

The determination of a non-Gaussian fixed  point for the theory of 
gravitation,  even in its basic Einstein Hilbert formulation, is an  essential 
step in the process of understanding not only  the (non-perturbative) 
renormalizability of the underlying  field theory but, more in general, the 
quantum nature of  gravity.  Although the presence of such a  fixed point 
(FP), originally  discovered in \cite{Weinberg:1976xy,Weinberg:1980gg} for 
gravity in  $d=2+\varepsilon$ dimensions,   was later confirmed  for  the  
$d=4$ dimensional case 
\cite{Reuter:1996cp,Falkenberg:1996bq,Lauscher:2001ya,Reuter:2001ag,Codello:2006in,Codello:2007bd,Codello:2008vh,Nink:2014yya,Gies:2015tca,DeBrito:2018hur}, a complete picture of the 
structure of the ultraviolet (UV)  critical manifold in $d=4$ still requires 
further  investigation. On general grounds, the approach  followed in 
\cite{Reuter:1996cp}  is based on the study of a Renormalization Group (RG) 
flow equation inspired by the original work of Wilson \cite{Wilson:1973jj}. 
Then, the presence  of a FP for the specific  flow equation considered, 
distinct from the Gaussian non-interacting  FP and characterized by a 
finite-dimensional UV critical surface (i.e. with a finite number of attractive  
eigendirections in the RG trajectory space),  guarantees that the theory is 
non-perturbatively  renormalizable and  indicated as asymptotically safe. In 
this context the gravitational theory is treated as a field theory where the 
elementary degrees of freedom are played by the components of the metric 
tensor and, in addition, we impose the physical requirement that the action 
which characterizes the theory must be invariant under diffeomorphisms of the 
metric.

Already in the definition of the basic modes of the metric $g_{\mu\nu}$, the 
system is not univocally constrained. In fact, after necessarily introducing a 
fiducial  background metric $\bg_{\mu\nu}$, one is allowed to treat the 
fluctuations $h_{\mu\nu}$ around  the background in different ways: in 
particular, the linear \cite{DeWitt:2003pm}
and the exponential \cite{Kawai:1993fq} are the most commonly adopted 
parametrizations and we shall consider both in the following.  Then, in the 
path integral formulation of the problem, the  diffeomorphism invariance of 
the functional integral can be implemented through the Faddeev-Popov method 
and this procedure brings in more freedom related to the selection of a 
particular gauge, which is essential to perform explicit calculations. 
Therefore, it is evident that both the freedom in the choice of the 
parametrization and of the gauge fixing function can lead to an unphysical 
dependence on some unconstrained parameters of various quantities extracted 
from the functional integration.

In principle one must distinguish between on-shell physical observables and 
off-shell quantities. In fact, while the former  must be invariant under any 
kind of symmetry transformation that leaves the physical system unmodified, 
there is no such restriction concerning the latter. So, for instance, while 
the  full effective action of a system is typically gauge dependent, physical 
quantities computed from the functional derivatives  of the effective action 
computed at its minimum, must not show any residual gauge dependence. However, 
the distinction between the two sets of variables can be spoiled  by other 
approximations that must unavoidably be introduced in the analysis of the 
problem. This includes the particular truncation of the action selected and 
also the approximation scheme adopted to explicitly solve the RG flow equation 
and therefore we can expect residual gauge and parametrization  dependence in 
the determination of any variable.

A thorough study of the residual gauge and parametrization  
dependence has already been carried out in \cite{Gies:2015tca}  by 
focusing on the non-Gaussian FP of the  Einstein-Hilbert action and 
on its UV critical  manifold. In particular, in \cite{Gies:2015tca} 
the RG flow equation for the  effective average action, introduced 
in \cite{Wetterich:1992yh,Morris:1993qb,Berges:2000ew}, is used to 
perform this analysis and, with the help of the principle of 
minimum sensitivity, a weak parametrization  dependence of the UV 
stable non-Gaussian FP is pointed out. However, as already noticed, 
even the details of the RG flow  equation play a crucial role in 
introducing  a redundant dependence on not  physically detectable
parameters, such as the dependence on the specific infrared 
regulator used to determine the flow \cite{Reuter:2001ag}.

Therefore, in this paper, we  analyze the same problem addressed in 
\cite{Gies:2015tca}, but instead of 
following the flow of the effective average (1PI) action, we focus 
on  a Wilsonian flow, closer in spirit to the 
original work of Wilson \cite{Wilson:1973jj}, subsequently 
developed in the work of Wegner and Houghton 
\cite{Wegner:1972ih}. More specifically, we shall refer to the 
proper time (PT) 
version of the flow equation 
\cite{Oleszczuk:1994st,Floreanini:1995aj,Liao:1994fp,Bohr:2000gp,
deAlwis:2017ysy}, which has been  employed in many contexts, from 
the renormalization  properties of scalar theories 
\cite{Bonanno:2004pq}, including the determination of critical 
exponents for the 3D Ising system 
\cite{Bonanno:2000yp,Mazza:2001bp,Litim:2010tt}, to the analysis of 
the  asymptotic safety of the conformal sector of quantum {gravity} 
\cite{Bonanno:2012dg,Bonanno:2023fij,Bonanno:2023ghc}, to  precise 
determination of observables in quantum mechanics,  of strictly non-perturbative nature \cite{Zappala:2001nv,Bonanno:2022edf}.

Although the PT equation does not belong to the class  of exact 
flows for the effective average action 
\cite{Litim:2001ky,Litim:2002xm,Wetterich:1992yh},  the possibility 
of interpreting the PT flow as a coarse grained  wilsonian flow has 
been recently been proposed by several authors 
\cite{deAlwis:2017ysy,Bonanno:2019ukb,
abel2023exactschwingerpropertime} and further discussed in \cite{wetterich2024simplifiedfunctionalflowequation}.

In fact the PT flow equation turns out to be extremely  accurate in 
the determination of observables  (even at first or second order 
truncation in the  derivative expansion), but it also has the 
feature of  preserving specific symmetries of the action, as in the 
case of the gauge symmetry  of Yang-Mills theories 
\cite{Liao:1995nm}, due to the gauge preserving nature of the 
proper time regulator introduced in \cite{Schwinger:1951nm}. 
Because of these properties, we want to investigate to which extent 
the residual dependence on unphysical  parameters coming from 
fluctuation parametrization and gauge fixing, in quantum gravity, 
does actually show up  in the PT flow.


In this respect, it must be noticed that  in a recent paper,  
\cite{Falls:2024noj}, the PT flow has been considered as the  
possible realization of a particular variant of the dimensional 
regularization which deals with poles appearing at all dimensions 
$d$, and also field redefinitions are  used to remove off-shell  
contributions to the RG equations, in the spirit of the essential 
renormalization group,
\cite{Falls:2017cze, Baldazzi:2021ydj},
where only the flow of  the  couplings which contribute to the 
scaling of physical observables is taken into account;  within 
this   approach the independence of the 
Newton's constant  beta function 
from  the parameterization is observed, 
to all orders in the  scalar curvature.

Here, we shall simply limit ourselves to keep track of   the  
persistence of  physically relevant features of the quantum 
Einstein gravity,  namely the presence of a non-Gaussian fixed 
point with positive values of the Newton and cosmological constant 
couplings and with two relevant, i.e. UV attractive, (real part of) 
eigenvalues that regulate the renormalizability of the theory, 
against changes in the parametrization of the fluctuations of the 
metric tensor and in the particular gauge choice. Concerning this 
latter issue we shall examine not only the usual renormalizable 
gauge \cite{Abbott:1980hw}, but also the results obtained by 
resorting to the physical gauge choice 
\cite{Percacci:2015wwa,Gaberdiel:2010xv,Zhang:2012kya,
Alexandre:2013iva,Percacci:2013ii,Labus:2015ska}.
In addition,  we cannot neglect, in any RG flow analysis, the 
freedom associated to the choice of  the particular regulator used 
to define the intermediate steps of the flow at different values of 
the energy scale $k$. For the PT flow, we introduce a commonly used 
parametrization of the regulator in terms of an integer parameter 
$m$, and we shall also consider two slightly different ways of  
implementing the regulator in the PT flow, according
to \cite{Liao:1994fp,Bonanno:2019ukb}.

Section II is dedicated to the definition of the problem together  with the introduction of all the unconstrained parameters entering  our analysis. Then, in Section III as a  simple test of our set-up, the parametrization dependence of the non-Gaussian FP in  $d=2+\varepsilon$ is considered, while Section IV is devoted to the  complete analysis of the model in $d=4$. Conclusions are reported in Section V while the Appendixes contain some relevant details of the complete calculations.

\section{Flow equation with different gauges and 
parametrizations}

The main goal of a quantum theory of Einstein gravity in a 
$d$-dimensional euclidean space-time is to evaluate the 
path integral 
\begin{equation}\label{generating functional gravity}
	Z=\frac{1}{V_{\text{diff.}}}\int D g_{\mu\nu} e^{-\mS[g]}
\end{equation}
over the metric $g_{\mu\nu}$ of a generic action $\mS[g]$ that is invariant under diffeomorphisms, with $V_{\text{diff.}}$ representing the redundancies of the symmetry.\\

The non-perturbative methods of the renormalization group are very useful for evaluating Eq. (\ref{generating functional gravity}). In this framework, our description of the theory is encoded in an effective action, which describes the theory as effective up to a momentum scale $\Lambda$. The same theory can be described at lower momentum scales using a flow equation that connects the actions at different scales.\\
In this work, we use a flow equation for the Wilsonian effective action $\mS_\Lambda$, regularized via the proper time integral \cite{deAlwis:2017ysy,Bonanno:2019ukb,Bonanno:2004sy}
 \begin{equation}
 	\Lambda \partial_\Lambda \mS_\Lambda=\frac12\int_0^\infty \frac{ds}{s} r_\Lambda (s) \STr e^{-s\tS_\Lambda^{(2)}}.
 	\label{PTRGE}
 \end{equation}
 Here, $\tS^{(2)}_\Lambda$ is the hessian of the gauge fixed Wilsonian effective action and of the ghost actions. The regularization is implemented through the regulating function the regulating function $\rho_{k,\Lambda}(s)$, which depends on the IR scale $k$ and the UV scale $\Lambda$, with
 \begin{equation}
 \label{regulator}
r_\Lambda(s)=\Lambda\partial_\Lambda\rho_{k,\Lambda}(s).
 \end{equation}
A standard approach to the problem concerns the use of the background field method, in which the metric $g_{\mu\nu}$ is decomposed into a classical background $\bar{g}_{\mu\nu}$ and a fluctuating metric $h_{\mu\nu}$ \cite{Abbott:1980hw}. This splitting can be carried out using different parameterizations, and in this work we are interested in the linear parametrization \cite{Reuter:1996cp}
\begin{equation}
    g_{\mu\nu}=\bg_{\mu\nu}+h_{\mu\nu},
\end{equation}
and in the exponential parametrization \cite{Nink:2014yya}
\begin{equation}
    g_{\mu\nu}=\bg_{\mu\rho}\left(e^h\right)^{\rho}_{\ \nu}.
\end{equation}
It is useful to summarize both parametrization through a dependence on the parameter $\tau$ \cite{Gies:2015tca}:
\begin{equation}
\label{tauparam}
	g_{\mu\nu}=\bg_{\mu\nu}+h_{\mu\nu}+\frac{\tau}{2}h_\mu^{\ \rho}h_{\rho \nu}+o(h^3),
\end{equation}
that for $\tau=0$ leads to the linear parameterization, while for $\tau=1$ it corresponds to the second-order expansion in $h_{\mu\nu}$ of the exponential parameterization.\\ In general, to distinguish between objects evaluated on the full metric and those evaluated on the background, we denote the latter as barred objects.\\We assume to have a maximally symmetric background metric, such that the Riemann and Ricci tensors are expressed solely in terms of the background metric and the curvature scalar $\bR$, \cite{Weinberg:1972kfs}
\begin{equation}
\bR_{\mu\nu\rho\sigma}=\frac{1}{d(d-1)}\left(\bg_{\mu\rho}\bg_{\nu\sigma}-\bg_{\mu\sigma}\bg_{\nu\rho}\right)\bR, \quad\quad\quad\quad\quad\quad \bR_{\mu\nu}=\frac{\bR}{d}\bg_{\mu\nu}.
\end{equation}
The fluctuation is decomposed with the York decomposition \cite{York:1973ia,Dou:1997fg}, and the path integral is rewritten in terms of the fields 
$h^T_{\mu\nu},\xi_\mu,\sigma,h$. A derivation of the York decomposition, together with the definitions of these fields 
is given in Appendix \ref{appendix York decomposition}.\\
With the freedom to fix the gauge in different ways, we adopt two different procedures.\\
As the first option, we implement the Faddeev-Popov gauge fixing procedure, and choose the background field gauge (BFG) \cite{Abbott:1980hw,Reuter:1996cp}
\begin{equation}                
F_\mu[h;\bg]=\left(\bg^{\alpha\gamma}\delta^\beta_\mu \overline D_\gamma-\omega\bg^{\alpha\beta}\overline D_\mu\right)h_{\alpha\beta},
\end{equation}
adding a gauge fixing term to the action
\begin{equation}
	\mS_{\text{gf}}[h;\bg]=\frac1\alpha \int d^dx \sqrt{\bar{g}} \bar{g}^{\mu\nu} F_\mu F_\nu,
\end{equation}
 and the Faddeev-Popov determinant in the usual form as an action for the ghosts $\bC^\mu$ and $C^\mu$
 \begin{equation}
 	\mS_{\text{gh}}[h,C,\bC;\bg]=-\int d^dx \sqrt{g} \bar{C}_\mu \frac{\delta F^\mu}{\delta \epsilon^\nu}C^\nu.
    \label{ghost action BFG}
 \end{equation}
Similarly to the fluctuation, the ghost field is also decomposed into its transverse and longitudinal components \cite{Dou:1997fg}:
\begin{equation}
C_\mu=\hat{C}_\mu+\bar{D}_\mu \frac{\eta}{\sqrt{{-\overline D^2}}}
 \end{equation}
with a trivial jacobian associated with $\{C_\mu\}\to\{\hat{C}_\mu,\eta\}$.\\
As we can see, this gauge fixing is parametrized by $\alpha$ and $\omega$. The gauge fixing is not well defined for the singular value 
\begin{equation}\label{singular omega}
    \omega_{\text{sing.}}=1,
\end{equation}
as observed in \cite{Gies:2015tca}.\\
The second option regards the possibility to fix some degrees of freedom of the theory, with the use of the physical gauge \cite{Percacci:2015wwa,Gaberdiel:2010xv,Zhang:2012kya,Alexandre:2013iva,Percacci:2013ii,Labus:2015ska}.\\
If we fix $\xi_\mu=0$, a ghost action appears, in term of the real ghost field $b_\mu$:
\begin{equation}
	\mS^{(\xi=0)}_{\text{gh}}=\int d^dx \sqrt{\bg} b^\mu \left(-\overline D^2-\frac{\bR}{d}+o(h)\right) b_\mu.
\end{equation}
We can also fix $\sigma=0$ or $h=0$. The associated ghost actions are, respectively in term of the scalar fields $\chi$ and $\phi$:
\begin{equation}
	\mS^{(\sigma=0)}_{\text{gh}}=\int d^dx \sqrt{\bg} \chi \left(-\overline D^2-\frac{\bR}{d-1}+o(h)\right) \chi,
\end{equation}
and
\begin{equation}
	\mS^{(h=0)}_{\text{gh}}=\int d^dx \sqrt{\bg} \phi \left(-\overline D^2+o(h)\right) \phi.
\end{equation}
The last two conditions cannot be imposed simultaneously; therefore, we use two possible gauge choices.\\
In the first case, we impose the conditions $\xi_\mu=0$ and $\sigma=0$, so the fluctuation is described only by $h^T_{\mu\nu}$ and $h$. We refer to this gauge choice as the ``T-h gauge" (ThG).\\
The second gauge is defined by $\xi_\mu=0$ and $h=0$. The action 
remains with $h^T_{\mu\nu}$ and $\sigma$, and we refer to this second 
gauge choice as the ``T-$\sigma$ gauge" (T$\sigma$G).\\The details 
concerning the gauge fixing procedures are shown in Appendix 
\ref{Appendix gauge fixing}.\\
The equation (\ref{PTRGE}) describes the flow of all possible couplings for $\mS_\Lambda[g]$. We take the so-called Einstein-Hilbert truncation, where the action is constrained to the form
 \begin{equation}
 	\mS_\Lambda=2\kappa^2Z_\Lambda\int d^dx \sqrt{g}\left(-R+2\bl_\Lambda\right),
 	\label{EH truncation}
 \end{equation}
 where $\kappa=(32\pi G)^{-1/2}$, with $G$ the Newton constant, and $Z_\Lambda=G/G_\Lambda$. This truncation is basically the Einstein-Hilbert action, with the Newton constant $G$ and the cosmological constant $\bl$ being ``promoted" to running coupling constants $G_\Lambda$ and $\bl_\Lambda$. Their values at low momentum scales must match the values of $G$ and $\bl$ in order to describe the physics of our universe.\\This ``promotion" is also reflected in the gauge fixing term, where we add an overall scale-dependent factor $\kappa^2 Z_\Lambda$.\\The scale dependence of the ghost sector for the Einstein-Hilbert truncation has been studied in \cite{Groh:2010ta}. One of the results of the work is that the wave-function renormalization of the ghosts exhibits an UV behavior similar to that of the graviton wave-function renormalization. Following this idea, we add $Z_\Lambda$ as an overall factor in the ghost actions.\\The hessian in the flow equation is evaluated in the background. In the BFG, expanding the gauge fixed action around the background, the quadratic term is
\begin{equation}
	\tS^{\text{quadratic}}_{\text{BFG,}\Lambda}=\frac12\int d^dx \sqrt{\bg}\left[h^T_{\mu\nu}\mS_{TT}h^{T\, \mu\nu}+\xi_\mu \mS_{\xi\xi}\xi^\mu+\begin{pmatrix}\sigma&h\end{pmatrix}\begin{pmatrix}\mS_{\sigma\sigma}&\mS_{h\sigma}\\ S_{h\sigma}&\mS_{hh}\end{pmatrix}\begin{pmatrix}\sigma\\h\end{pmatrix}\right],
\end{equation}
while the ghost action, being already quadratic in the ghost fields, is evaluated in the background metric
\begin{equation}
	\mS^{\text{gh}}_{\text{BFG}}=\int d^dx \sqrt{\bg}\left[\bar{\hat{C}}_\mu \mS_{CC} {\hat{C}}^\mu+\bar{\eta}\mS_{\eta\eta}\eta\right].
\end{equation}
This hessian is almost diagonal, and we exploit the possibility to fix 
\begin{equation}\label{constraint}
    \omega=\frac{2+d\alpha-2\alpha}{2d},
\end{equation}
value for which $\mS_{h\sigma}=0$, therefore the hessian is diagonal. With this approach our strategy differs from that adopted in \cite{Gies:2015tca}. The flow equation \eqref{PTRGE} requires the exponentiation of the hessian and in our approach it is a trivial step. Moreover, each term of the hessian has the form
\begin{equation}
	\mS_{ii}=-A_{\Lambda,i}\overline D^2+B_{\Lambda,i}\bl_\Lambda+C_{\Lambda,i} \bR,
    \label{hessian form}
\end{equation}
for $i\in \Omega_{BFG}\equiv\{T,\xi,\sigma,h,C,\eta\}$.
With the constraint \eqref{constraint}, the singular value \eqref{singular omega} is reflected in
\begin{equation}
\label{alfacrit}
    \alpha_{\text{sing.}}=2\frac{d-1}{d-2}
\end{equation}
the coefficients $A_{\Lambda,\sigma},A_{\Lambda,h},A_{\Lambda,\eta}$ are null and this yields a singularity in the flow equation.\\
In the ThG, the quadratic part of the action, expanded around the background, is
\begin{equation}
    \tS^{\text{quadratic}}_{\text{ThG,}\Lambda}=\frac12\int d^dx \sqrt{\bg}\left[h^T_{\mu\nu}\mS_{\tilde{T}\tilde{T}}h^{T\,\mu\nu}+h\mS_{\tilde{h}\tilde{h}}h\right],
\end{equation}
and the ghost action
\begin{equation}
    \tS^{\text{gh}}_{\text{ThG,}\Lambda}=\frac12\int d^dx \sqrt{\bg}\left[b_\mu\mS_{bb}b^\mu+\chi\mS_{\chi\chi}\chi\right].
\end{equation}
Alternatively, in the T$\sigma$G, we find
\begin{equation}
    \tS^{\text{quadratic}}_{\text{T$\sigma$G,}\Lambda}=\frac12\int d^dx \sqrt{\bg}\left[h^T_{\mu\nu}\mS_{\tilde{T}\tilde{T}}h^{T\,\mu\nu}+\sigma\mS_{\tilde{\sigma}\tilde{\sigma}}\sigma\right],
\end{equation}
and
\begin{equation}
    \tS^{\text{gh}}_{\text{T$\sigma$G,}\Lambda}=\frac12\int d^dx \sqrt{\bg}\left[b_\mu\mS_{bb}b^\mu+\phi\mS_{\phi\phi}\phi\right].
\end{equation}
With the physical gauges, the hessian is automatically diagonal, and each term of the hessian takes the form (\ref{hessian form}), with $i\in\Omega_{ThG}\equiv\{\tilde{T},\tilde{h},b,\chi\}$ in the ThG, $ i\in\Omega_{T\sigma G}\equiv\{\tilde{T},\tilde{\sigma},b,\phi\}$ in the T$\sigma$G.\\The explicit form of the components of the hessians is reported in the Appendix \ref{appendix components of the hessian}.\\
With the hessians decomposed into pieces that can be written in the form in Eq. (\ref{hessian form}), the regularization is carried out straightforwardly. The function $\rho_{k,\Lambda}(s)$ 
defined in Eq. \eqref{regulator}, provides an IR and an UV regularization, as it vanishes both for $s>1/k^2$ and  $s<1/\Lambda^2$. We shall test the flow equation for different 
families of regulators, also parametrized by the integer $m>d/2$, 
as previously discussed in \cite{Bonanno:2019ukb}.
Namely, by taking 
\begin{equation}
\label{reg1}
\rho_{k,\Lambda}(s;m)=\frac{\Gamma\left(m,msk^2\right)-\Gamma\left(m,ms\Lambda^2\right)}{\Gamma\left(m\right)},
\end{equation}
and $r_\Lambda(s;m)$ according to Eq. \eqref{regulator},
\begin{equation}
\label{reg2}
    r_\Lambda(s;m)=\frac{2}{\Gamma\left( m\right)}\left(ms\Lambda^2\right)^m e^{-ms\Lambda^2} \; ,
\end{equation}
the exponential term suppresses modes with  $A_{\Lambda,i} p^2>1/s$. Then, since we want an effective suppression of the modes with $p^2>1/s$, we have to suitably rescale the regularization functions in Eqs. \eqref{reg1}, \eqref{reg2}, by the factor $A_{\Lambda,i}$. Therefore we take
\begin{equation}
\rho_{k,\Lambda}(A_{\Lambda,i}s;m)=\frac{\Gamma\left(m,mA_{\Lambda,i}sk^2\right)-\Gamma\left(m,mA_{\Lambda,i}s\Lambda^2\right)}{\Gamma\left(m\right)}, 
\label{rhoB}
\end{equation}
and
\begin{equation}
	r_\Lambda(A_{\Lambda,i} s;m)=\frac{2}{\Gamma(m)}\left(1+\frac{\Lambda}{2}\partial_\Lambda \log A_{\Lambda,i}\right)\left(mA_{\Lambda,i} s\Lambda^2\right)^m e^{-mA_{\Lambda,i} s\Lambda^2}.
	\label{b scheme}
\end{equation}
In the following, we refer to this regularization as the ``B  scheme".
Instead, if we insert the factors $A_{\Lambda,i}$ directly in the derivative, we have
\begin{equation}
	r_\Lambda(A_{\Lambda,i} s;m)=\frac{2}{\Gamma(m)}\left(mA_{\Lambda,i} s\Lambda^2\right)^m e^{-mA_{\Lambda,i} s\Lambda^2}\; ,
	\label{c scheme}
\end{equation}
and we indicate the latter as the ``C scheme".
For the B scheme, Eq. (\ref{PTRGE}) becomes
\begin{equation}
		\Lambda \partial_\Lambda \mS_\Lambda[g]=\left(1+\frac{\Lambda}{2}\partial_\Lambda\log Z_{\Lambda}\right)\sum_{i}\STr_i\left(\frac{m\Lambda^2A_{\Lambda,i}}{\mS_{ii}+m\Lambda^2A_{\Lambda,i}}\right)^m,
		\label{PTRGE Scheme B}
\end{equation}
while for  the C scheme we get
\begin{equation}
	\Lambda \partial_\Lambda \mS_\Lambda[g]=\sum_{i}\STr_i\left(\frac{m\Lambda^2A_{\Lambda,i}}{\mS_{ii}+m\Lambda^2A_{\Lambda,i}}\right)^m\;,
	\label{PTRGE Scheme C}
\end{equation}
where $\STr_i$ is the supertrace over the degrees of freedom of the 
$i$-th field. In our case, the supertrace is simply the trace with 
a factor of $-2$ for the complex ghosts and a factor of $-1$ for 
the real ghosts.\\

Eq. \eqref{PTRGE Scheme B} differs from eq. \eqref{PTRGE Scheme C}
because of the factor 
$\frac{\Lambda}{2}\partial_\Lambda\log Z_{\Lambda}$.
The former is a direct consequence of the truncation choice. With 
$Z_\Lambda$ in the ghost sector, each term of the hessian has the 
same structure, and the factor $\Lambda \partial_\Lambda\log 
Z_\Lambda$ appears overall. In contrast, in the latter scheme, the 
presence of $Z_\Lambda$ in the ghost sector is not necessary to get 
the flow equation \eqref{PTRGE Scheme C}.
It must be remarked that the C scheme should not be considered 
as a further approximation of the B scheme where some (but not all)
dependence on the wave function renormalization is neglected. 
Instead, it could be  regarded as a different scheme, not produced 
by Eq. \eqref{rhoB}, but rather from another $\rho_{k,\Lambda}$
whose derivative gives Eq. \eqref{PTRGE Scheme C}.

The traces in the two flow equations are over the fields defined by 
the 
York decomposition and they can be handled by using the Heat Kernel 
expansion, as shown in \cite{Lauscher:2001ya}. In the right-hand 
side 
of these equations, the flow of $\bl_\Lambda/G_\Lambda$ is the 
coefficient of $(1/8\pi)\int d^dx \sqrt{g}$, and the flow of  
$1/G_\Lambda$ is the coefficient of $(-1/16\pi)\int d^dx 
\sqrt{g}R$. 
It is convenient to write the expressions in terms of the 
dimensionless quantities $g_\Lambda=\Lambda^{d-2}G_\Lambda$, 
$\lambda_\Lambda=\Lambda^{-2}\bl_\Lambda$. In this way, the two beta 
functions for the dimensionless coupling constants are:
\begin{align}
	\label{betag}\Lambda\partial_\Lambda g_\Lambda=&\beta_g\left(g_\Lambda,\lambda_\Lambda;d,\text{gauge},\alpha,\text{parametrization},\text{scheme},m\right)\equiv (d-2+\eta_N)g_\Lambda\\
	\label{betalambda}\Lambda\partial_\Lambda \lambda_\Lambda=&\beta_\lambda\left(g_\Lambda,\lambda_\Lambda;d,\text{gauge},\alpha,\text{parametrization},\text{scheme},m\right)\equiv \left(\eta_N-2\right)\lambda_\Lambda+\Phi
\end{align}
where $\eta_N\equiv -\Lambda\partial_\Lambda \log Z_\Lambda$. The beta functions show a dependence on the dimension, the gauge choice, the parametrization, and the regulating scheme, and they can be written in terms of $\eta_N$ and $\Phi$. Their values are shown in Appendix \ref{appendice eta phi} for the scenarios discussed below: in the BFG, ThG or T$\sigma$G; with the linear or exponential parametrization
(respectively $\tau=0$ or $\tau=1$ in Eq. \eqref{tauparam} ); 
and finally  with the C or B regulating scheme.

\section{Running of $g_\Lambda$ in $d=2+\varepsilon$}
We start by checking our set-up, and more specifically 
Eq. (\ref{betag}) with $\lambda_\Lambda=0$ fixed,
against the well-settled case of the
fixed point in  $d=2+\varepsilon$, (see \cite{Weinberg:1976xy,Weinberg:1980gg}). 
By performing  an expansion around $g=0$ and $\varepsilon=0$, we get:
\begin{equation}
	\beta_g=(\varepsilon+\eta_N)g_\Lambda\equiv\varepsilon g_\Lambda+ Ag_\Lambda^2+o(\varepsilon g_\Lambda)+o(\varepsilon^2)+o(g^2_\Lambda) \;.
\end{equation}
The value of $A$ can be computed analytically and it turns out to be
independent on the regulating scheme, yet it shows a dependence on 
the gauge and parametrization choices. For the BFG and the ThG we find $A=-\frac{38}{3}-4\tau$, while for the T$\sigma$G, $A=-\frac{26}{3}-4\tau$.  A curious remark is that the beta function  has the same value for the first two gauges in the linear parametrization 
($\tau=0$) and for the last gauge in the exponential parametrization ($\tau=1$).\\
It is clear that the different results are due to the hessian of the effective action. In fact, the terms that make it up show a sensitivity to the parametrization, and furthermore, different terms appear when different gauges are used.\\
A deeper analysis of this result consists in comparing the various contributions, obtained by splitting the flow equation into pieces generated by the different terms in the hessian as, for instance, the traceless-transverse contributions in the ThG and C scheme:
\begin{equation}
	\text{traceless-transverse contribution}\equiv\STr_{\tilde{T}}\left(\frac{m\Lambda^2A_{\Lambda,\tilde{T}}}{\mS_{\tilde{T}\tilde{T}}+m\Lambda^2A_{\Lambda,\tilde{T}}}\right)^m\;, 
\end{equation}
and the analogous contributions coming from the scalar $h$, from the 
vector ghost and from the scalar ghost. Their sum gives back the full 
flow. The same procedure is applied for for each scenario and the 
results are separately reported in  Table \ref{tabella1}. 
\renewcommand{\arraystretch}{2.1}
\begin{table}[ht]
    \centering
    \begin{tabular}{|c|c|c|c|}
    \hline
	&BFG&ThG&T$\sigma$G\\
	\hline
	traceless-transverse & $-12$ & $-12$ & $-12$\\
	transverse vector & $\dfrac23$ &  & \\
	scalar $\sigma$ & $\dfrac{14}{3}$& &$\dfrac{14}{3}-4\tau$ \\
	scalar $h$ & $\dfrac{14}{3}-4\tau$& $\dfrac{14}{3}-4\tau$& \\
	transverse vector ghost & $-\dfrac43$& $-\dfrac23$& $-\dfrac23$\\
	scalar ghost & $-\dfrac{28}{3}$ &$-\dfrac{14}{3}$ & $-\dfrac23$\\	
	\hline
    total & $-\dfrac{38}{3}-4\tau$ &$-\dfrac{38}{3}-4\tau$ & $-\dfrac{26}{3}-4\tau$\\	
	\hline
    \end{tabular}
    \caption{Contribution to  A in Eq. \eqref{hessian form}  from  different sectors of the hessian, for the three different gauge choices.}
    \label{tabella1}
\end{table}
\renewcommand{\arraystretch}{1}
In the data of Table \ref{tabella1},
we find no dependence  either on the parameter $m$ or on the scheme 
(B or C), while they are dependent on  $\tau$ and on the gauge 
choice.\\
We notice that the dependence on the parametrization 
is entirely carried by the scalar parts in all gauges. 
Then, if we compare the BFG and the ThG, we see that the 
differences arise in the transverse vector and scalar parts, 
including the ghosts. 
But, as shown in 
Table \ref{tabella2}, if we take the sum of the transverse vector 
and the transverse vector ghost part,  the result in the two gauges is the same, and the 
same  happens for the scalar and the scalar ghost part.
\renewcommand{\arraystretch}{2.1}
\begin{table}[ht]
    \centering
    \begin{tabular}{|c|c|c|c|}
    \hline
	&BFG&ThG&T$\sigma$G\\
	\hline
	spin-2 & $-12$ & $-12$ & $-12$\\
	spin-1 & $-\dfrac23$ & $-\dfrac23$ & $-\dfrac23$\\
	spin-0 & $-4\tau$& $-4\tau$ &$4-4\tau$ \\
	\hline
    total & $-\dfrac{38}{3}-4\tau$ &$-\dfrac{38}{3}-4\tau$ & $-\dfrac{26}{3}-4\tau$\\	
	\hline
    \end{tabular}
    \caption{Contribution to  A in Eq. \eqref{hessian form}  from  different sectors of the hessian, grouped by their spin.}
    \label{tabella2}
\end{table}
\renewcommand{\arraystretch}{1}
The different value of $A$ for the T$\sigma$G, when compared to the ThG, is entirely due to the spin-0 part, with a difference of $4$, the same difference between the results for the linear and exponential parametrizations.
\section{non-Gaussian fixed point in $d=4$}
We are now able to search for the non-Gaussian fixed point (NGFP) 
$(\lambda_\Lambda^*,g_\Lambda^*)\neq(0,0)$ of the two beta functions 
(\ref{betag}), (\ref{betalambda}), and analyze its stability in $d=4$ 
for the various scenarios introduced above and therefore, in addition 
to the regulation scheme (B or C) and to the specific gauge choice, 
we must take into account the dependence on the parameter $m$ in the 
regulator function, and also
(for the BFG gauge only) on the gauge fixing parameter $\alpha$.
In the following we shall select three representative values of $m$,
namely $m=3,4,8$; in fact the first two  values are the smallest 
integers allowed for this parameter, according to the 
constraint $m>d/2$, and $m=8$ essentially corresponds to 
the value which provides the best determination of the 
universal index $\nu$ for the critical 3D Ising model, 
at least to  the lowest order truncation of the derivative expansion
\cite{Litim:2010tt}.\\
The gauge fixing requires non-negative values of $\alpha$. Nevertheless, we investigate the dependence on this parameter in the interval $(-10,10)$, by analytically continuing the expressions to negative $\alpha$.
\subsection*{Background field gauge}
\begin{figure}[ht]
    \centering
    \includegraphics[width=0.7\linewidth]{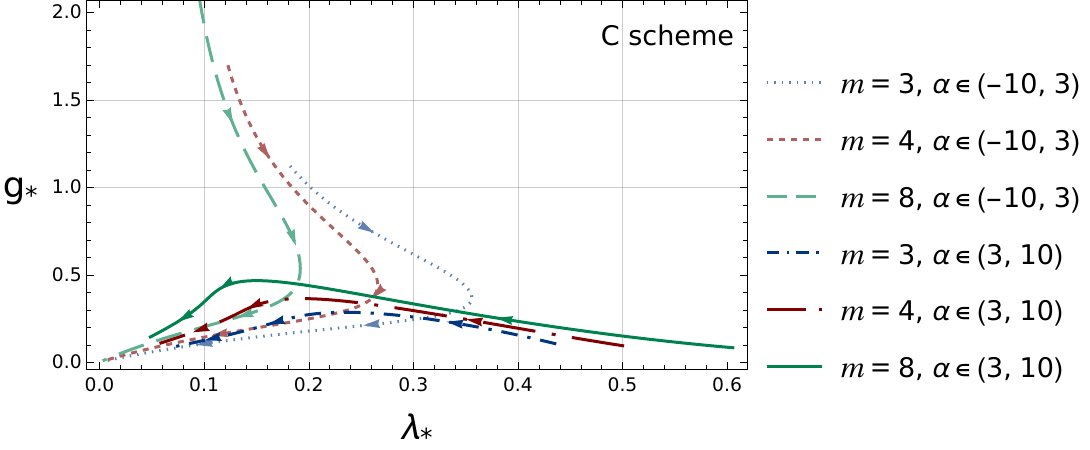}
    \includegraphics[width=0.7\linewidth]{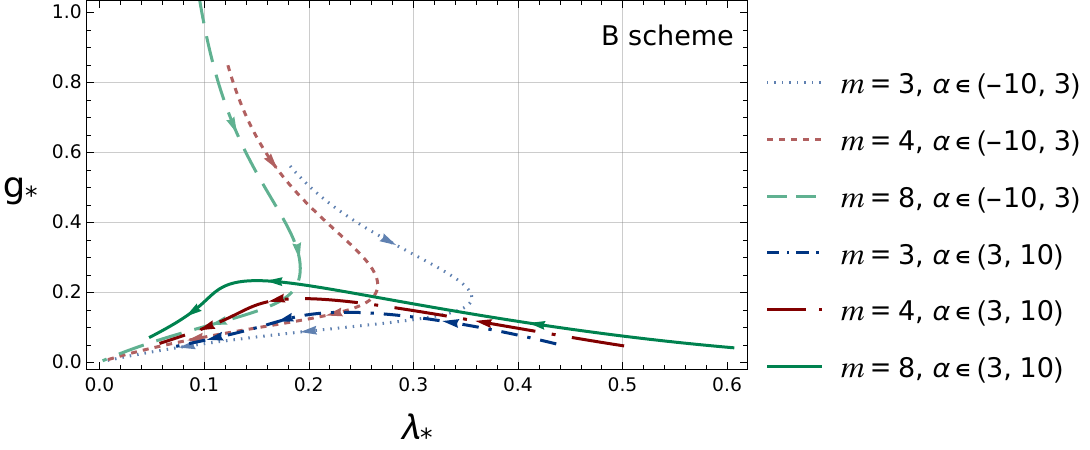}
    \caption{NGFP in the BFG and linear parametrization for the two regularization schemes B (lower panel) and C (upper panel). 
    Each line corresponds to a value of $m$ and each dot on a line corresponds to one  particular value of $\alpha$. $\alpha$ increases, following the arrow, from $-10$ to $3$ below the singularity $\alpha=3$ and it grows from $3$ to $10$ above the singularity.}
    \label{fpBFGLinear}
\end{figure}

\begin{figure}
    \centering
    \includegraphics[width=0.7\linewidth]{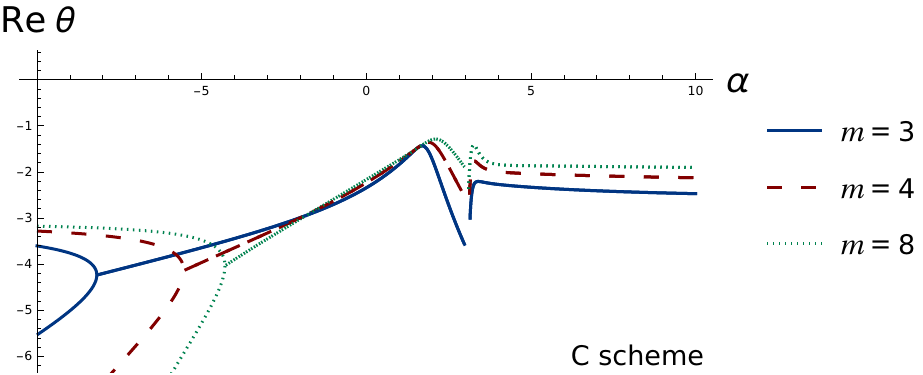}
    \includegraphics[width=0.7\linewidth]{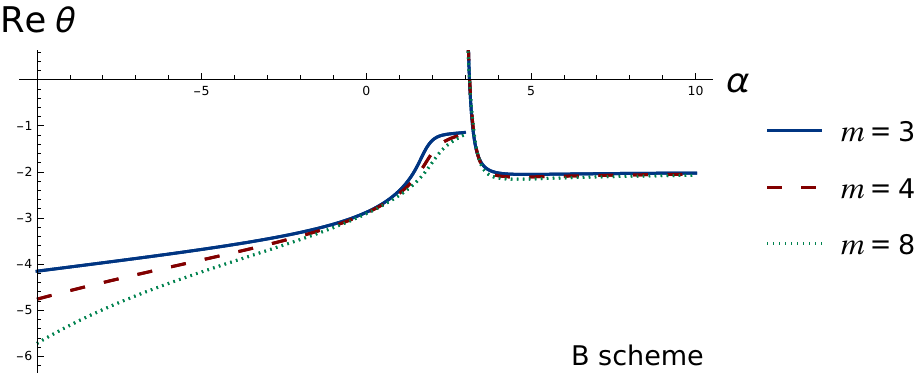}
    \caption{The real part of the eigenvalues $\theta$ of the stability matrix evaluated at the NGFP with the BFG and the linear parametrization for the two regularization schemes, B (lower 
    panel) and C (upper panel).  As in Fig. \ref{fpBFGLinear},
    each line is parameterized the gauge fixing coefficient $\alpha$  and corresponds to a particular value of $m$. 
    The two plots show the effects of the singularity at $\alpha=3$, and for the C scheme, the presence of a bifurcation, illustrating how the eigenvalues change from real to complex.}
    \label{ReBFGLinear}
\end{figure}
Let us begin by analyzing the scenario with the BFG and linear 
parametrization ($\tau=0$). Both in the B and C schemes, we find the 
NGFP for different values of $m$ and $\alpha$ with the exception of 
$\alpha=3$ (obtained from  Eq. \eqref{alfacrit} for $d=4$) 
where, as mentioned above, the flow shows a singularity. 
The fixed point position on the plane depends on the scenarios and the 
parameters but it is always located in the first quadrant as
shown in Fig. \ref{fpBFGLinear}.

Concerning the stability issue, we 
examine the real part of the eigenvalues $\theta$ of the stability 
matrix. Away from the singular point, Re$\,\theta$ is negative, 
independently of the  parameters and the scheme. Thus, the fixed point 
is UV attractive, as shown in  Fig. \ref{ReBFGLinear} 
(note that our convention on the sign of the critical exponents
is opposite to the one adopted in \cite{Gies:2015tca}). 
The eigenvalues in 
the B scheme are always a pair of complex conjugates, while in the C scheme they bifurcate for negative $\alpha$ and become a pair of real eigenvalues. In proximity of the singularity, the stability of the fixed point strongly depends on the scheme. In fact, in the B scheme, the NGFP is attractive when the singularity is approached from the left, while it becomes repulsive when it is approached from the right. In contrast, in the C scheme, the NGFP remains attractive in both regions.\\

\begin{figure}
    \centering
    \includegraphics[width=0.7\linewidth]{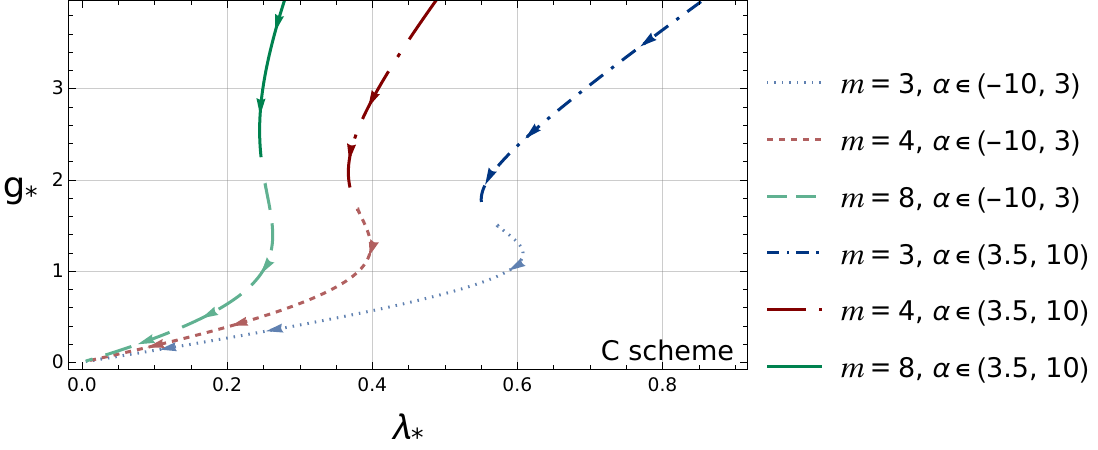}
    \includegraphics[width=0.7\linewidth]{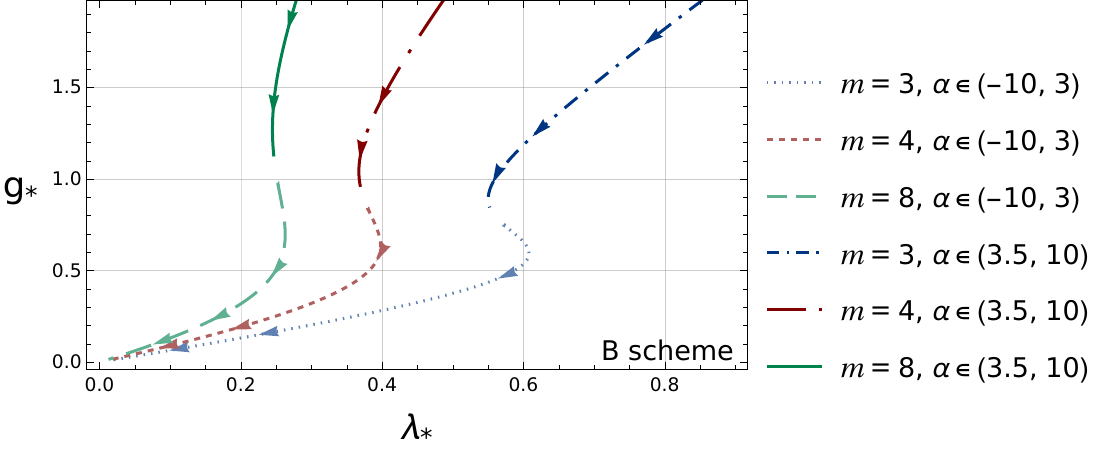}
    \caption{NGFP with the BFG and exponential parametrization for the two regulating schemes, B (lower panel) and C (upper panel). Coding is as in Fig. \ref{fpBFGLinear}. Values of $\alpha$ within interval $3<\alpha \lesssim 3.5$ are not plotted as in this region the NGFP solution is not found. }
    \label{fpBFGExponential}
\end{figure}

\begin{figure}
    \centering
    \includegraphics[width=0.7\linewidth]{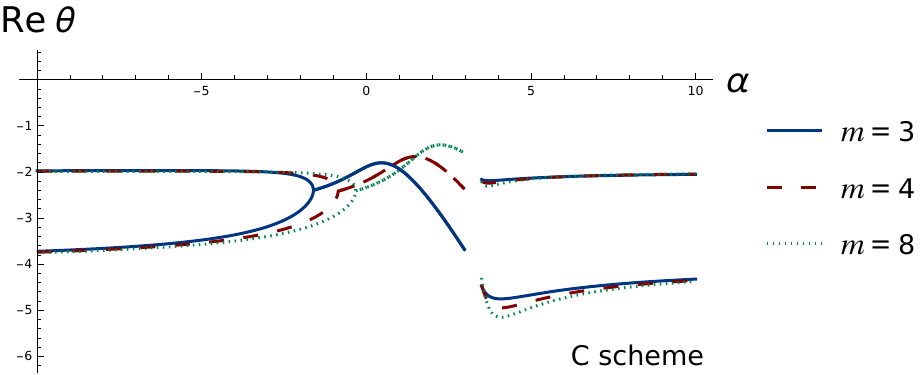}
    \includegraphics[width=0.7\linewidth]{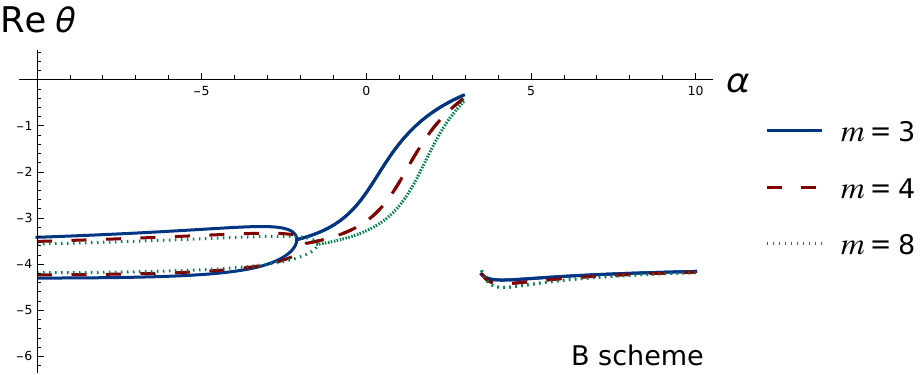}
    \caption{Real part of the eigenvalues $\theta$ of the stability matrix evaluated at the NGFP, with the BFG and exponential parametrization, for the two regulating schemes, B (lower panel) and C (upper panel). Coding is as in Fig. \ref{ReBFGLinear}.
    As in the case of the upper panel of Fig. \ref{ReBFGLinear},
    the presence of bifurcations, corresponding to the change of the eigenvalues from real to complex, is observed.}
    \label{ReBFGExponential}
\end{figure}

   By switching to the scenario in the BFG and exponential parametrization, we find the NGFP for different values of $m$ and $\alpha$ in both schemes, with the exception of the region close  to the singularity, where we find the NGFP on the left of the singularity while on the right side it disappears in the interval $3<\alpha\lesssim3.5$. The fixed point is located in the first quadrant and its location depends both on the scheme and the parameters (see Fig. \ref{fpBFGExponential}).\\ The NGFP (when it exists) is UV attractive independently of the scenario. The dependence shows up only in the values of the eigenvalues as illustrated in Figure \ref{ReBFGExponential}. With a bifurcation, they go from real to complex and, for $\alpha>3.5$, they are real in the C scheme and complex in the B scheme.

\begin{table}[ht]
\begin{minipage}[l]{8.4truecm}
	\[
	\begin{tabular}{ |c|c|c|c|c|c|  }
		\hline
		\multicolumn{6}{|c|}{ThG - linear parametrization} \\
		\hline
		scheme& $m$ & $\lambda_\Lambda^*$ &$g_\Lambda^*$&$\text{Re} \theta$&$\text{Im}\theta$\\
		\hline
		C & $3$ & 0.3310 & $ 0.5152$ & $ -2.343$ & $ 1.325 $\\
		C & $4$ & $ 0.2474$ & $ 0.6303$ & $ -2.234$ & $ 1.148$\\
		C & $8$ & $ 0.1792$ & $ 0.7813$ & $ -2.165$ & $ 1.009$\\
		\hline
		B & $3$ & $ 0.3310$ & $ 0.2576$ & $ -2.875$ & $ 2.494$\\
		B & $4$ & $ 0.2474$ & $ 0.3152$ & $ -2.891$ & $ 2.063$\\
		B & $8$ & $ 0.1792$ & $ 0.3907$ & $ -2.904$ & $ 1.727$\\
		\hline
	\end{tabular}
	\]
\end{minipage}
\begin{minipage}[r]{8.4truecm}
	\[
		\begin{tabular}{ |c|c|c|c|c|c|  }
			\hline
			\multicolumn{6}{|c|}{ThG - exponential parametrization} \\
			\hline
			scheme& $m$ & $\lambda_\Lambda^*$ &$g_\Lambda^*$&$\text{Re} \theta$&$\text{Im}\theta$\\
			\hline
			C & $3$ & $0.5992$ & $ 1.066$ & $ -1.889$ & $ 1.326$ \\
			C & $4$ & $ 0.3975$ & $ 1.3140$ & $ -2.190$ & $ 0.686$\\
			C & $8$ & $ 0.2610$ & $ 1.5790$ & $ -2.336$ & $ 0.359$\\
			\hline
			B & $3$ & $ 0.5992$ & $ 0.5330$ & $ -2.447$ & $ 2.161$\\
			B & $4$ & $ 0.3975$ & $0.6570 $ & $ -3.064$ & $ 1.072$\\
			B & $8$ & $ 0.2610$ & $ 0.7895$ & $ -3.276$ & $ 0.665$\\
			\hline
		\end{tabular}
	\]
\end{minipage}
\begin{minipage}[l]{8.4truecm}
	\[
		\begin{tabular}{ |c|c|c|c|c|c|  }
			\hline
			\multicolumn{6}{|c|}{T$\sigma$G - linear parametrization} \\
			\hline
			scheme& $m$ & $\lambda_\Lambda^*$ &$g_\Lambda^*$&$\text{Re} \theta$&$\text{Im}\theta$\\
			\hline
			C & $3$ & $0.3133$ & $ 0.5783$ & $ -2.567$ & $ 1.345$ \\
			C & $4$ & $ 0.2325$ & $ 0.7161$ & $ -2.477$ & $ 1.198$\\
			C & $8$ & $ 0.1677$ & $ 0.8975$ & $ -2.424$ & $ 1.086$\\
			\hline
			B & $3$ & $ 0.3133$ & $ 0.2892$ & $ -3.035$ & $ 2.754$\\
			B & $4$ & $ 0.2325$ & $0.3580 $ & $ -3.061$ & $ 2.402$\\
			B & $8$ & $ 0.1677$ & $ 0.4487$ & $ -3.084$ & $ 2.145$\\
			\hline
		\end{tabular}
	\]
\end{minipage}
\begin{minipage}[r]{8.4truecm}
	\[
		\begin{tabular}{ |c|c|c|c|c|c|  }
			\hline
			\multicolumn{6}{|c|}{T$\sigma$G - exponential parametrization} \\
			\hline
			scheme& $m$ & $\lambda_\Lambda^*$ &$g_\Lambda^*$&$\phantom{\text{R}}\theta_1\phantom{\text{e}}$&$\phantom{\text{R}}\theta_2\phantom{\text{e}}$\\
			\hline
			C & $3$ & $0.5625$ & $ 1.5708$ & $ -4$ & $ -2$ \\
			C & $4$ & $ 0.375$ & $ 1.76715$ & $ -4$ & $ -2$\\
			C & $8$ & $ 0.25$ & $ 2.06167$ & $ -4$ & $ -2$\\
			\hline
			B & $3$ & $ 0.5625$ & $ 0.785398$ & $ -4$ & $ -4$\\
			B & $4$ & $ 0.375$ & $0.883573 $ & $ -4$ & $ -4$\\
			B & $8$ & $ 0.25$ & $ 1.03084$ & $ -4$ & $ -4$\\
			\hline
		\end{tabular}
	\]
\end{minipage}
    \caption{NGFP and critical exponents in the physical gauge.}
    \label{tabella3}
    
\end{table}

\subsection*{Physical gauge}

Within the physical gauge, there is no singularity in the 
flow, and we 
find the NGFP in the first quadrant and, as before, its 
location depends on the specific scenario. The eigenvalues 
are complex in the 
ThG, while in the T$\sigma$G, the eigenvalues are complex for the 
linear parametrization but they are real for the 
exponential one. 
In all scenarios, the NGFP is UV attractive, as shown in 
Table 
\ref{tabella3}.\\

For the T$\sigma$G in the exponential parametrization, the 
eigenvalues are independent of the parameter $m$. To understand why 
this scenario generate this property, we must return to the (\ref{TT 
physical gauge}), (\ref{sigmsigma physical gauge}), (\ref{b physical 
gauge}) and (\ref{phi phi physical gauge}), which, for $\tau=1$, are 
independent of $\bl_\Lambda$, and $Z_\Lambda$ appears only as multiplicative factor. After the integration over $s$, in the C scheme, it is clear that the RHS of the flow equation does not depend of the coupling constants. Thus the flow equation is simply
\begin{equation} \label{flow simplified 1}
    \Lambda \partial_\Lambda \mS_\Lambda=\int d^dx \sqrt{g}\left(\Omega_0(d,m) \Lambda^d+\Omega_1(d,m) R \
\Lambda^{d-2}+o(R^2)\right),
\end{equation}
where we included all the dependencies of the regulator in the coefficient $\Omega_0(d,m)$ and $\Omega_1(d,m)$.\\
Straightforwardly the beta functions are \eqref{betag}, \eqref{betalambda}, with $\eta_N=16\pi \Omega_1(d,m) g_\Lambda$ and $\Phi=8\pi \Omega_0(d,m) g_\Lambda$.\\
These beta functions are quadratic in the couplings, making it possible to analytically determine the NGFP:
 \begin{equation}
    \left(\lambda_\Lambda^*,g_\Lambda^*\right)=\left(\dfrac{\Omega_0(d,m)}
    {\Omega_1(d,m)}\frac{2-d}{2d},\dfrac{2-d}{16\pi \Omega_1(d,m)}\right).
     \label{NGFP simplified 1}
 \end{equation}
 Clearly, the NGFP depends on the regulating parameter $m$, while for the 
 critical exponents, i.e. the eigenvalues of the stability matrix,  this dependence cancels out, with the result
 \begin{equation}\label{critical exponents simplified 1}
     \theta_1=-d, \quad \theta_2=2-d.
 \end{equation}
 In the B scheme the flow equation is the same with only a multiplicative factor overall
\begin{equation}\label{flow simplified 2}
    \Lambda \partial_\Lambda \mS_\Lambda=\int d^dx \sqrt{g}\left(\Omega_0(d,m) \Lambda^d+\Omega_1(d,m) R \
\Lambda^{d-2}+o(R^2)\right)\left(1-\frac{\eta}{2}\right)
\end{equation}
and the beta functions are the previous with the prescriptions \eqref{etaBscheme}, \eqref{PhiBscheme}.\\
The NGFP is
 \begin{equation}     
\left(\lambda_\Lambda^*,g_\Lambda^*\right)=\left(\dfrac{\Omega_0(d,m)}
    {\Omega_1(d,m)}\frac{2-d}{2d},\dfrac{2-d}{8\pi d \Omega_1(d,m)}\right).
 \label{NGFP simplified 2}
\end{equation}
It shows dependence on $m$, whereas the critical exponents are
 \begin{equation}\label{critical exponents simplified 2}
     \theta_1=-d, \quad \theta_2=\frac12 d(2-d).
 \end{equation}
So, in both schemes the position of the fixed point is not 
universal, while the critical exponents are independent of $m$. We 
verified that the second eigenvalue is different in the two schemes 
due to the presence of the factor $\left(1-\frac{\eta}{2}\right)$ 
in the B scheme that originates from the factor $\frac{\Lambda}
{2}\partial_\Lambda\log Z_\Lambda$ in Eq. \eqref{PTRGE Scheme B}, 
unlike in the C scheme where such terms are absent, as noticed 
before.\\

The same results can also be obtained within a different approach,
i.e. by using the renormalization group flow equation for the effective average action $\Gamma_k$ 
\cite{Wetterich:1992yh,Morris:1993qb,Berges:2000ew}
     \begin{equation}
	\partial_t \Gamma_k=\frac12 \STr\left[\frac{\partial_t R_k}{\Gamma^{(2)}_k+R_k}\right]=\frac12 \sum_i\STr\left[\frac{\partial_t R_{k,i}}{\Gamma_{ii}+R_{k,i}}\right] \;,
	\label{Wetterich}
\end{equation}
where, in the last step, the same notation used for the Wilsonian effective action is adopted, with $\Gamma_{ii}$ identical to $\mS_{ii}$. $R_{k,i}$ is the regulator that can be written as $R_{k,i}=A_{k,i}k^2 R^{(0)}\left(\frac{-\overline D^2}{k^2}\right)$, with $R^{(0)}\left(z\right)$ an appropriate regulating function. Then,
\begin{equation}
	\partial_t R_{k,i}=\partial_t A_{k,i} k^2 R^{(0)}\left(\frac{-\overline D^2}{k^2}\right)+2A_{k,i} k^2 R^{(0)}\left(\frac{-\overline D^2}{k^2}\right)-2 A_{k,i} k^2\left(\frac{-\overline D^2}{k^2}\right) \left[R^{(0)}\left(\frac{-\overline D^2}{k^2}\right)\right]'
    \label{R derivative}
\end{equation}
and, if we neglect the $\partial_t A_{k,i}$ in Eq. 
(\ref{R derivative}), $A_{k,i}$ is an overall factor in $\partial_t R_{k,i}$, thus the RHS of the Eq. (\ref{Wetterich}) is independent of 
the coupling constants. This leads us directly to the Eq. \eqref{flow simplified 1}, where the two coefficients now depend of the regulating function instead of $m$.\\ The NGFP is then the \eqref{NGFP simplified 1} and the critical exponents are the \eqref{critical exponents simplified 1}.\\

On the other hand, if we neglect $\left[R^{(0)}\left(\frac{-\overline 
D^2}{k^2}\right)\right]'$ in Eq. (\ref{R derivative}), it can be written as
\begin{equation}
	\partial_t R_{k,i}=2A_{k,i}k^2 R^{(0)}\left(\frac{-\overline D^2}{k^2}\right)\left(1-\frac{\eta}{2}\right),
\end{equation}
then the Eq. (\ref{Wetterich}) behaves as the Eq. \eqref{flow simplified 2}, thus the NGFP is the \eqref{NGFP simplified 2} and the critical exponents are the \eqref{critical exponents simplified 2}
\section{Conclusions}

The non-perturbative renormalizability of QEG  is a direct 
consequence of the existence of a non-gaussian FP, characterized by 
a finite dimensional UV attractive critical  manifold for the RG 
flow of the associated  running action. Therefore, the goal of the 
present analysis is to provide further evidence  to the already  
existing research, that residual dependence on non-physical 
parametrizations, necessary to define the flow, disappear
(at least partially) when one focuses on the physically 
relevant properties of such a critical manifold. 

In fact, it is known that,  for a  truncated effective action of a diffeomorphism-invariant theory of gravity,
quantities such as beta functions depend not only on the  details of  the  regularization but also on field 
parametrization and gauge fixing  choice.  However,  in a 
physically meaningful picture, it is desirable that the residual 
unphysical dependence, does actually disappear when establishing 
the existence  of a non-trivial solution  of the FP equation (we 
recall here that its specific location  - unlike its existence -  
is not  an universal property) and  the sign  of the eigenvalues  
of the linearized flow in proximity of the FP.

To this aim,  in our analysis, we selected the Einstein-Hilbert truncation  and used either the linear or exponential parametrization, the background field gauge or the physical gauge and,  when resorting  to the PT  flow equation, we retained  two suitable families of parametrized regulators.  With this set-up , the search for non-perturbative FP in $d=2+\varepsilon$ dimensions, shows that such a solution is always guaranteed but its specific value shows a the residual parametrization  dependence, which is entirely carried by the spin-0 degrees of freedom, while the spin-2 and spin-1 components remain independent.
Furthermore, the results for the Newton and cosmological  constants 
in $d=4$ dimensions confirm the presence of  such residual  
dependence.

More specifically, 
when resorting to  the background gauge, 
even though  the (real part of the) eigenvalues plotted 
in Figs. \ref{ReBFGLinear} and 
\ref{ReBFGExponential} show, at least qualitatively, a weak 
dependence on the regulator parameter $m$ (see for instance the 
bifurcation of the eigenvalue solution occurring
for negative $\alpha$ in all plots, with the exception of 
the lower panel of Fig. \ref{ReBFGLinear}), the dependence on the
regulator scheme (B or C) and on the linear or exponential parametrization is significant.
On the other hand, the dependence on the gauge parameter $\alpha$,
quantitatively evident for $\alpha\leq 4$,  practically 
vanishes for larger $\alpha$.

When switching to the physical gauge, the values of the  real part 
of the eigenvectors in Table \ref{tabella3} indicate a mild dependence both 
on the parametrization and on the parameter $m$, 
and a stronger dependence on the scheme B or C, with
$|\text{Re}\,\theta|$ slightly larger in the former case.
In this framework,  it is  noteworthy that  the absence 
of the cosmological constant  in the hessian, obtained  with 
exponential parametrization 
and with ``T-$\sigma$  gauge",  directly implies  $m$-parameter  
independent eigenvalues. 

Finally a comparison between the results in the 
two gauges is not particularly significant, due to the large numbers of 
parameters involved, and one can only notice that the changes in $\text{Re}\,\theta$
are reasonably limited, as long as the parameter $\alpha$ does not approach 
the singular value $\alpha=3$.
In any case, for almost all  configurations analyzed,  
independently of the adopted field parametrization, gauge 
choice and regulator  parametrization,  we always find a UV-
attractive non-Gaussian fixed point  with positive 
values of the coupling constants,thus confirming an asymptotically 
safe scenario even for the PT flow.

As already mentioned, the gauge and parametrization  dependence 
has already been analyzed  in \cite{Gies:2015tca}. 
When confronting   the 
approach followed in this work  with ours,  it is necessary to remark the difference both in the RG  flow equations  adopted and in the implementation of the background field 
gauge fixing.  In particular, they do not impose the constraint 
\eqref{constraint} and investigate the dependence on  $\beta=d\omega-1$ for  fixed values of 
$\alpha$. Thus, a direct comparison between the two results is not  possible, but it is nevertheless interesting that these  different approaches reach the same conclusion about the  the robustness of the asymptotic safety hypothesis and, in addition, the critical exponents in the two  cases share 
similar properties, such as bifurcations from real to  complex values. Finally, it would be interesting to 
investigate whether a suitable restriction to the flow of essential operators only, according to the 
spirit  of  \cite{Falls:2024noj}, could further suppress  the residual unphysical parametrization dependence
in the beta functions of the theory.

\acknowledgements{We would like to thank K. Falls and R. Ferrero for useful comments on the manuscript. A.B. would like to thank Antonio Puglisi for substantial help in an early investigation of this problem and Frank Saueressig for discussions. G.O. would like to thank Emiliano M. Glaviano and Gian Paolo Vacca for fruitful discussions.}

\begin{appendices}
\section{York decomposition}\label{appendix York decomposition}
The fluctuating metric, being a symmetric 2-tensor, can be 
decomposed into a spin-2, a spin-1 and two spin-0 components. In 
fact, we can separate it into its traceless and trace parts
\begin{equation}
    h_{\mu\nu}=h^{\text{Traceless}}_{\mu\nu}+\frac1d\bg_{\mu\nu}h,
\end{equation}
with $h=\bg_{\mu\nu}h^{\mu\nu}$ the first spin-0 component, and 
$\bg^{\mu\nu}h^{\text{Traceless}}_{\mu\nu}=0$. 
From the traceless part, we isolate its transverse component 
$h^T_{\mu\nu}$, which satisfies the condition $\overline D^\mu 
h^T_{\mu\nu} = 0$, while the remaining degrees of freedom are 
described by a vector $v^\mu$:
\begin{equation}
    h^{\text{Traceless}}_{\mu\nu}=h^T_{\mu\nu}+\overline D_\mu v_\nu+\overline D_\nu v_\nu-\frac2d\bg_{\mu\nu} \overline D^\lambda v_\lambda.
\end{equation}
$h^T_{\mu\nu}$ is the spin-2 tensor, while $v_\mu$ is decomposed into a transverse and a longitudinal part, respectively the spin-1 and spin-0 components:
\begin{equation}
    v_\mu=\hx_\mu+\overline D_\mu \frac{\hs}{2},
\end{equation}
with $\overline D^\mu \hx_\mu=0$. In this way we arrive at the York decomposition:
\begin{equation}
	h_{\mu\nu}=h^T_{\mu\nu}+\bar{D}_\mu  \hx_\nu +\bar{D}_\nu \hx_\mu+\left(\bar{D}_\mu \bar{D}_\nu-\frac1d \bar{g}_{\mu\nu}\bar{D}^2\right) \hs+\frac1d \bar{g}_{\mu\nu} h.
\end{equation}
This decomposition carries a non-trivial jacobian in the path integral. We rescale
\begin{equation}
	\hx_\mu=\left(-\bar{D}^2-\frac{\bar{R}}{d}\right)^{-\frac12}\xi_\mu, \quad \text{ and } \quad \hs=\left(-\bar{D}^2\right)^{-\frac12}\left(-\bar{D}^2-\frac{\bar{R}}{d-1}\right)^{-\frac12}\sigma,
\end{equation}
and the path integral in terms of $\{h^T_{\mu\nu},\xi_\mu,\sigma,h\}$ has a trivial jacobian.
\section{Gauge fixing in QEG}\label{Appendix gauge fixing}
An infinitesimal diffeomorphism $\epsilon$ is described by the Lie derivative of the metric:
\begin{equation}
	\delta g_{\mu\nu}=L_\epsilon g_{\mu\nu}\equiv\epsilon^\rho \partial_\rho g_{\mu\nu}+g_{\mu\rho}\partial_\nu\epsilon^\rho+g_{\rho\nu}\partial_\mu \epsilon^\rho=D_\mu\epsilon_\nu+D_\nu\epsilon_\mu,
	\label{infinitesimal diffeomorphism for the metric}
\end{equation}
under which the action $\mS_\Lambda[g]$ remains invariant.\\

With the metric split into background and fluctuation components, we 
must define their respective transformations. One possibility, dubbed  
background gauge transformation, the background metric transforms as
\begin{equation}
\delta_B \bg_{\mu\nu}=L_\epsilon \bg_{\mu\nu}
\label{background gauge transformations for background}
\end{equation}
thus, both for the linear and the exponential parametrization
\begin{equation}
	\delta_B h_{\mu\nu}=L_\epsilon h_{\mu\nu}.
	\label{background gauge transformations for fluctuation}
\end{equation}
Another option, indicated as quantum gauge transformation,  
corresponds to no background transformation:
\begin{equation}
	\delta_Q \bg_{\mu\nu}=0\; ,
	\label{quantum gauge transformations for background}
\end{equation}
and therefore,  in the linear parametrization, we have
\begin{equation}
    \delta_Q h_{\mu\nu}=L_\epsilon g_{\mu\nu},
	\label{quantum gauge transformations for fluctuation linear 
    parametrization}
\end{equation}
while with the exponential parametrization:
\begin{equation}
	\bg_{\mu\rho}\delta_Q \left(e^ h\right)^\rho_{\ \nu}=L_\epsilon 
    g_{\mu\nu}.
	\label{quantum gauge transformations for fluctuation exponential 
    parametrization}
\end{equation}
For our purpose, after the expansion of Eq. \eqref{quantum gauge
transformations for fluctuation exponential parametrization} in powers 
of $h^\rho_{\ \nu}$, we find
\begin{equation}
    \bg_{\mu\rho}\left(e^h\right)^\rho_{\ \alpha}\left(\delta_Q 
    h^\alpha_{\ \nu}+\frac12 \left[\delta_Q h,h\right]^\alpha_{\ \nu} 
    \right)+o(h^2)=	L_\epsilon g_{\mu\nu}.
\end{equation}
Then, since we are interested in $\delta_Q h_{\mu\nu}=\delta_Q\left(g_{\mu\gamma}h^\gamma_{\ \nu}\right)$, the transformation with the exponential parametrization is
\begin{equation}
    \delta_Q h_{\mu\nu}=L_\epsilon g_{\alpha\beta}\left(\delta^\alpha_\mu\delta^\beta_\nu+\frac12\delta^\alpha_\mu h^\beta_{\ \nu}+\frac12 h^\alpha_{\ \mu} \delta^\beta_\nu\right)+o(h^2)
	\label{delta quantum h mu nu exponential}
\end{equation}
The quantum transformations for the fluctuation are clearly different for the two parameterizations, but they yield the same value when $h=0$. Then, for both parameterizations, the transformation can be expressed as
\begin{equation}
    \delta_Q h_{\mu\nu}=L_\epsilon \bg_{\mu\nu}+o(h).
	\label{delta quantum h mu nu}
\end{equation}

\subsection{Background Field Gauge}
We fix the gauge by using the quantum gauge transformations, imposing the condition
\begin{equation}
	F_\mu[h;\bg]=\left(\bg^{\alpha\gamma}\delta^\beta_\mu \overline D_\gamma-\omega\bg^{\alpha\beta}\overline D_\mu\right)h_{\alpha\beta}.
    \label{gauge condition}
\end{equation}
With the Faddeev-Popov procedure, we remove the redundancies in the path integral by inserting the gauge fixing action 
\begin{equation}
	\mS_{\text{gf}}[h;\bg]=\frac{1}{\alpha}\int d^dx \sqrt{\bg} \bg^{\mu\nu} F_\mu F_\nu,
\end{equation}
and the Faddeev-Popov determinant $\det\left(\frac{\delta F_\mu}{\delta \epsilon^\nu}\right)$, which is written as a path integral over the ghost fields $C_\mu, \bC_\mu$ of the action
\begin{equation}
	\mS_{\text{gh}}[h,C,\bC;\bg]=-\int d^dx \sqrt{\bg}\bC_\mu \frac{\delta F^\mu}{\delta \epsilon^\nu}C^\nu.
\end{equation}
Then, Eq. (\ref{generating functional gravity}) is now
\begin{equation}
    Z=\int Dh^T_{\mu\nu} D\xi_\mu D\sigma Dh DC_\mu D\bC_\mu e^{-\mS[g]-\mS_{\text{gf}}[h;\bg]-\mS_{\text{gh}}[h,C,\bC;\bg]}.
    \end{equation}
This gauge fixing procedure exhibits a dependence on the 
parametrization. While the (\ref{gauge condition}) has been chosen 
independently of it, the Faddeev-Popov determinant, which depends on 
the quantum gauge transformations, differs in  
the two parametrizations.\\
In fact, in the linear parametrization
\begin{equation}
    \frac{\delta F_\mu}{\delta \epsilon^\nu}=\left(\bg^{\alpha\gamma}\delta^\beta_\mu \overline D_\gamma-\omega\bg^{\alpha\beta}\overline D_\mu\right)\left(g_{\nu\beta}D_\alpha+g_{\nu\alpha}D_\beta\right),
\end{equation}
while in the exponential parametrization
\begin{equation}
    \frac{\delta F_\mu}{\delta \epsilon^\nu}=\left(\bg^{\alpha\gamma}\delta^\beta_\mu \overline D_\gamma-\omega\bg^{\alpha\beta}\overline D_\mu\right)\left(g_{\nu\delta}D_\rho+g_{\nu\rho}D_\delta\right)\left(\delta^\rho_\alpha\delta^\delta_\beta+\frac12\delta^\rho_\alpha h^\delta_{\ \beta}+\frac12 h^\rho_{\ \alpha} \delta^\delta_\alpha\right)+o(h^2)
\end{equation}
In the flow equation, we are interested in the truncation at $h=0$; thus, the ghost action results the same in the two parametrizations. 

\subsection{Physical Gauge}
The idea of the physical gauge is to strongly constrain the redundant degrees of freedom in the path integral.\\
If we decompose the vector generating the infinitesimal diffeomorphism into its transverse and longitudinal components
\begin{equation}
	\epsilon_\mu=\epsilon^T_\mu+\overline D_\mu \frac{\psi}{\sqrt{-\overline D^2}}, \quad\quad \text{with }\overline D^\mu\epsilon^T_\mu=0,
\end{equation}
Eq. (\ref{delta quantum h mu nu}) is
\begin{equation}
    \delta h_{\mu\nu}=\overline D_\mu \epsilon^T_\nu+\overline D_\nu \epsilon^T_\mu+\overline D_\mu\overline D_\nu \frac{2\psi}{\sqrt{-\overline D^2}}+o(h).
\end{equation}
This transformation must be implemented  
in the many fields of the York decomposition
\begin{equation}
	\delta h_{\mu\nu}=\delta h^T_{\mu\nu}+\left(\bar{D}_\mu \delta \hx_\nu +\bar{D}_\nu \delta \hx_\mu\right)+\left(\bar{D}_\mu \bar{D}_\nu-\frac1d \bar{g}_{\mu\nu}\bar{D}^2\right)\delta \hs+\frac1d \bar{g}_{\mu\nu}\delta h.
\end{equation}
The trace of the two equations gives the transformation
\begin{equation}
	\delta h=-2\sqrt{-\overline D^2}\psi+o(h).
\end{equation}
We  isolate the transverse and longitudinal components of the transformation
\begin{equation}
	\delta\hs=\frac{2\psi}{\sqrt{-\overline D^2}}+o(h),
\end{equation}
\begin{equation}
	\delta\hx_\mu=\epsilon^T_\mu+o(h),
\end{equation}
\begin{equation}
	\delta h^T_{\mu\nu}=o(h),
\end{equation}
and write the transformations of $\hs$ and $\hx_\mu$ in term of the rescaled fields $\sigma$ and $\xi_\mu$:
\begin{equation}
	\delta\sigma=2\sqrt{-\overline D^2-\frac{\bR}{d-1}}\psi+o(h),
\end{equation}
\begin{equation}
	\delta\xi_\mu=\sqrt{-\overline D^2-\frac{\bR}{d}}\epsilon^T_\mu+o(h).
\end{equation}
Up to $o(h)$ terms, $h^T_{\mu\nu}$ is invariant, $\xi_\mu$ transforms with $\epsilon^T_\mu$, while $\sigma$ and $h$ transform with $\psi$. We can use these transformations to fix the gauge. $\epsilon^T$ can be fixed to have $\xi_\mu=0$, yielding a determinant in the path integral, which can be written as an integration over the real ghost field $b_\mu$, a transverse vector:
\begin{equation}
	\mS^{(\xi=0)}_{\text{gh}}=\int d^dx \sqrt{\bg} b^\mu \left(-\overline D^2-\frac{\bR}{d}+o(h)\right) b_\mu.
\end{equation}
Similarly, $\psi$ can be fixed to have $h=0$ or $\sigma=0$. Here as well, there is a determinant, which we write as an integration over the real ghost fields $\phi$ and $\chi$, respectively:
\begin{equation}
	\mS^{(h=0)}_{\text{gh}}=\int d^dx \sqrt{\bg} \phi \left(-\overline D^2+o(h)\right) \phi,
\end{equation}
and
\begin{equation}
	\mS^{(\sigma=0)}_{\text{gh}}=\int d^dx \sqrt{\bg} \chi \left(-\overline D^2-\frac{\bR}{d-1}+o(h)\right) \chi.
\end{equation}
If we impose the conditions $\xi_\mu=0$ and $\sigma=0$, the fluctuation is described only by $h^T_{\mu\nu}$ and $h$. We refer to this gauge choice as the ``T-h gauge" (ThG):
\begin{equation}
    Z=\int D h^{T}_{\mu\nu} Dh D b_\mu D\chi  e^{-\left.\mS[g]\right|_{\xi_\mu=\sigma=0}-\mS^{(\xi=0)}_{\text{gh}}-\mS^{(\sigma=0)}_{\text{gh}}}.
\end{equation}
If we impose the conditions $\xi_\mu=0$ and $h=0$, the fluctuation is described only by $h^T_{\mu\nu}$ and $\sigma$. We refer to this gauge choice as the ``T-$\sigma$ gauge" (T$\sigma$G):
\begin{equation}
    Z=\int D h^{T}_{\mu\nu} D\sigma D b_\mu D\phi  e^{-\left.\mS[g]\right|_{\xi_\mu=h=0}-\mS^{(\xi=0)}_{\text{gh}}-\mS^{(h=0)}_{\text{gh}}}.
\end{equation}
For the linear and the exponential parametrizations, the ghost actions differ in the $o(h)$ term. When we consider the truncation in our flow equation, $o(h)$  vanishes, and the two parametrizations yield the same ghost actions.

\section{Components of the hessian}\label{appendix components of the hessian}
\subsection{Background field gauge}
The traceless-transverse part is
\begin{equation}
	\mS_{TT}=2\kappa^2 Z_\Lambda\left[-\frac12 \overline D^2+\left(\tau-1\right)\bl_\Lambda+\frac{d^2-3d+4-\tau\left(d^2-3d+2\right)}{2d(d-1)}\bR\right]
\end{equation}
the transverse vector parts are
\begin{align}
	\mS_{\xi\xi}=&2\kappa^2 Z_\Lambda\left[-\frac{1}{\alpha}\overline D^2+\left(\tau-1\right)2\bl_\Lambda+\frac{\alpha(d-2)-1-\tau\alpha(d-2)}{\alpha d}\bR\right]\\
	\mS_{CC}=&Z_\Lambda\left[-\overline D^2-\frac{\bR}{d}\right]
\end{align}
and the scalar parts are
\begin{align}
	\mS_{\sigma\sigma}=&2\kappa^2 Z_\Lambda\left[\frac{(d-1)\left(2+d(\alpha-2)-2\alpha\right)}{2\alpha d^2} \overline D^2+\frac{d-1}{d}(\tau-1)\bl_\Lambda+\frac{\alpha(1-\tau)(d^2-3d+2)+2-2d}{2\alpha d^2}\bR\right]\\
	\mS_{h\sigma}=&\kappa^2 Z_\Lambda\frac{(1-d)(\alpha(d-2)-2d\omega+2)}{\alpha d^2}\sqrt{-\overline D^2\left(-\overline D^2-\frac{\bR}{d-1}\right)}\\
	\mS_{hh}=&2\kappa^2 Z_\Lambda\left[\frac{\alpha(d^2-3d+2)-2(d\omega-1)^2}{2\alpha d^2}\overline D^2+\frac{d+2(\tau-1)}{2d}\bl_\Lambda+\frac{-d^2+6d-8-2\tau(d-2)}{4d^2}\bR\right]\\
	\mS_{\eta\eta}=&2Z_\Lambda(1-\omega)\left[-\overline D^2-\frac{\bR}{d(1-\omega)}\right]
\end{align}
\subsection{Physical gauge}
For the ThG, in the hessian we find the traceless-transverse part
\begin{equation}
	\mS_{\tilde{T}\tilde{T}}=2\kappa^2 Z_\Lambda\left[-\frac{\overline D^2}{2}+ (\tau -1) \bl_\Lambda+\frac{2-(\tau-1)\left(d^2-3 d+2\right)}{2 d (d-1)}\bR\right],
	\label{TT physical gauge}
\end{equation}
the scalar part
\begin{equation}
	\mS_{\tilde{h}\tilde{h}}=2\kappa^2 Z_\Lambda\left[\frac{(d-1) (d-2)}{2 d^2}\overline D^2+\frac{d+2(\tau -1)}{2d}\bl_\Lambda-\frac{(d-2) (d+2 \tau -4)}{4 d^2}\bR\right],
\end{equation}
the ghost transverse vector part
\begin{equation}
	\mS_{bb}=2Z_\Lambda\left[-\overline D^2-\frac{\bR}{d}\right],
	\label{b physical gauge}
\end{equation}
and the ghost scalar part
\begin{equation}
	\mS_{\chi\chi}=2Z_\Lambda\left[-\overline D^2-\frac{\bR}{d-1}\right].
\end{equation}
Differently, for the T$\sigma$G the hessian is composed by traceless-transverse part (\ref{TT physical gauge}), the scalar part
\begin{equation}
	\mS_{\tilde{\sigma}\tilde{\sigma}}=2\kappa^2 Z_\Lambda\left[\frac{(d-1) (d-2) }{2 d^2}\overline D^2+(\tau -1)\frac{d-1}{ d}\bl_\Lambda-(\tau -1)\frac{d^2-3 d+2}{2 d^2}\bR\right],
    \label{sigmsigma physical gauge}
\end{equation}
the transverse ghost part (\ref{b physical gauge}), and the scalar ghost part
\begin{equation}
	\mS_{\phi\phi}=-2Z_\Lambda\overline D^2.
    \label{phi phi physical gauge}
\end{equation}

\section{$\eta_N$ and $\Phi$ for various scenarios}
\label{appendice eta phi}

We start by displaying $\eta_N$ and $\Phi$ in the 
BFG with  linear parametrization and C scheme of regularization:
\begin{align}\nonumber
	\eta^{\text{lin,bfg}}_N=&2^{2-d} \pi ^{1-\frac{d}{2}}  \frac{\Gamma \left(-\frac{d}{2}+m+1\right)}{{3 \Gamma (m)}} \left\{\frac{2 m^m}{d} \left[-12 (\alpha +1)+(1-6 \alpha ) d^2+(18 \alpha +5) d\right] (m-2 \alpha  \lambda_\Lambda )^{\frac{d}{2}-m-1}\right.\\
	& \nonumber
	\left.-\frac{d+1}{d} \left(5 d^2-22 d+48\right) m^m (m-2 \lambda_\Lambda )^{\frac{d}{2}-m-1}+\left(\frac{48}{-2 \alpha +(\alpha -2) d+2}-4 d+\frac{48}{d}-24\right) m^{\frac{d}{2}-1}\right.\\
	&\nonumber
	\left.+\frac{2 m^m \left[(\alpha +4) d-2 (\alpha +11)\right]}{2 d \lambda_\Lambda +m [-2 \alpha +(\alpha -2) d+2]}\left(\frac{2 d \lambda_\Lambda }{-2 \alpha +(\alpha -2) d+2}+m\right)^{\frac{d}{2}-m}\right.\\
	&\left.+\frac{2 m^m [(7 \alpha -2) d-2 (7 \alpha +5)]}{2 \alpha  d \lambda_\Lambda +m [-2 \alpha +(\alpha -2) d+2]}\left(\frac{2 \alpha  d \lambda_\Lambda }{-2 \alpha +(\alpha -2) d+2}+m\right)^{\frac{d}{2}-m}\right\}g_\Lambda \;,
\end{align}

\begin{align}
	\nonumber
	\Phi^{\text{lin,bfg}}=&-\frac{\Gamma \left(m-\frac{d}{2}\right)}{\Gamma (m)}2^{2-d} \pi ^{1-\frac{d}{2}}  \left\{4 d m^{d/2}-2 m^m \left(\frac{2 d \lambda_\Lambda }{-2 \alpha +(\alpha -2) d+2}+m\right)^{\frac{d}{2}-m}\right.\\
	& \nonumber
	-2 m^m \left(\frac{2 \alpha  d \lambda_\Lambda}{-2 \alpha +(\alpha -2) d+2}+m\right)^{\frac{d}{2}-m}-2 (d-1) m^m (m-2 \alpha  \lambda_\Lambda)^{\frac{d}{2}-m}\\
	& \biggl.
	-(d-2) (d+1) m^m \left(m-2 \lambda_\Lambda\right)^{\frac{d}{2}-m}\biggr\}g_\Lambda \; .
\end{align}
Then, in the BFG with exponential parametrization and C scheme of regularization, $\eta_N$ and $\Phi$ are:
\begin{align}\nonumber
	\eta_N^{\text{exp,bfg}}=&2^{2-d} \pi ^{1-\frac{d}{2}} \frac{\Gamma \left(-\frac{d}{2}+m+1\right)}{3 \Gamma (m)} \left\{\left[d \left(d^2+d-16\right)-48\right] \frac{m^{\frac{d}{2}-1}}{d}\right.\\ \nonumber
	&+\frac{2 (d-2) \left[(\alpha +4) d-2 (\alpha +5)\right]m^m}{2 d^2 \lambda_\Lambda +(d-2) \left[-2 \alpha +(\alpha -2) d+2\right]m}\left(\frac{2 d^2 \lambda_\Lambda }{(d-2) [-2 \alpha +(\alpha -2) d+2]}+m\right)^{\frac{d}{2}-m}\\ 
	&\left.+\frac{[2 (\alpha -2) d-4 (\alpha +5)] m^{\frac{d}{2}-1}}{-2 \alpha +(\alpha -2) d+2}+\left(\frac{48}{-2 \alpha +(\alpha -2) d+2}-4 d+\frac{48}{d}-24\right) m^{\frac{d}{2}-1}\right\}g_\Lambda
    \;,
\end{align}
\begin{align}
	\nonumber\Phi^{\text{exp,bfg}}=&2^{2-d} \pi ^{1-\frac{d}{2}} \frac{\Gamma \left(m-\frac{d}{2}\right)}{\Gamma (m)} \biggl\{\left(d^2-3 d-2\right) m^{d/2}\biggr.
	\\
	&\left.+2 m^m \left(\frac{2 d^2 \lambda_\Lambda }{(d-2) \left[-2 \alpha +(\alpha -2) d+2\right]}+m\right)^{\frac{d}{2}-m}\right\}g_\Lambda \;.
\end{align}
In the ThG with linear parametrization 
and C scheme of regularization, $\eta_N$ and $\Phi$ are:
\begin{align}\nonumber
	\eta_N^{\text{lin,ThG}}=&2^{2-d}\pi^{1-\frac{d}{2}}\frac{\Gamma\left(-\frac{d}{2}+m+1\right)}{3\Gamma(m)}\left\{\frac{12-12d+5d^2+d^3}{d(1-d)}2m^{\frac{d}{2}-1}\right.\\
	&\left.-\frac{1+d}{d}\left(48-22d+5d^2\right)m^m\left(m-2\lambda_\Lambda\right)^{\frac{d}{2}-m-1}+2\frac{2 d-11}{1-d} m^m \left(m-\frac{d \lambda_\Lambda }{d-1}\right)^{\frac{d}{2}-m-1}\right\}g_\Lambda \;,
\end{align}
\begin{align}\nonumber
	\Phi^{\text{lin,ThG}}=&2^{2-d} \pi ^{1-\frac{d}{2}} \frac{\Gamma \left(m-\frac{d}{2}\right)}{\Gamma (m)} \biggl\{-2 d m^{\frac{d}{2}}+2 m^m \left(m-\frac{d \lambda_\Lambda }{d-1}\right)^{\frac{d}{2}-m}
	\\ & +(d-2) (d+1) m^m (m-2 \lambda_\Lambda )^{\frac{d}{2}-m}\biggr\}g_\Lambda \; .
\end{align}
In the ThG with exponential parametrization and C scheme of regularization, $\eta_N$ and $\Phi$ are:
\begin{align}\nonumber
	\eta_N^{\text{exp,ThG}}=&2^{2-d}\pi^{1-\frac{d}{2}}\frac{\Gamma\left(-\frac{d}{2}+m+1\right)}{3\Gamma(m)}\biggl\{\frac{\left(d^3-4 d^2-35 d+26\right) }{d-1}m^{\frac{d}{2}-1}
	\\&-2\frac{2 d-5}{d-1}m^m \left(m-\frac{d^2 \lambda }{d^2-3 d+2}\right)^{\frac{d}{2}-m-1}\biggr\}g_\Lambda \; ,
\end{align}
\begin{align}\nonumber
	\Phi^{\text{exp,ThG}}=&2^{2-d} \pi ^{1-\frac{d}{2}} 
    \frac{\Gamma \left(m-\frac{d}{2}\right)}{\Gamma (m)} \biggl
    \{\left(d^2-3 d-2\right) m^{d/2}+2 m^m \left(m-\frac{d^2 
    \lambda }{d^2-3 d+2}\right)^{\frac{d}{2}-
    m}\biggr\}g_\Lambda \;.
\end{align}
In the T$\sigma$G with linear parametrization  and C scheme of regularization, $\eta_N$ and $\Phi$  are:
\begin{align}\nonumber
	\eta_N^{\text{lin,T$\sigma$G}}=&-2^{2-d}\pi^{1-\frac{d}{2}}\frac{\Gamma\left(-\frac{d}{2}+m+1\right)}{3\Gamma(m)}\biggl\{\frac{2}{d}\left(d^2+6 d-12\right) m^{\frac{d}{2}-1}-14 m^m \left(\frac{2 d \lambda_\Lambda }{d-2}+m\right)^{\frac{d}{2}-m-1}\\
	&+\frac{1+d}{d}\left(5 d^2-22 d+48\right) m^m (m-2 \lambda_\Lambda )^{\frac{d}{2}-m-1}\biggr\} g_\Lambda \; ,
\end{align}
\begin{align}\nonumber
	\Phi^{\text{lin,T$\sigma$G}}=&2^{2-d} \pi ^{1-\frac{d}{2}} \frac{\Gamma \left(m-\frac{d}{2}\right)}{\Gamma (m)} \biggl\{-2 d m^{d/2}+2 m^m \left(\frac{2 d \lambda_\Lambda }{d-2}+m\right)^{\frac{d}{2}-m}
	\\&+(d-2) (d+1) m^m (m-2 \lambda_\Lambda )^{\frac{d}{2}-m}\biggr\}g_\Lambda \; ,
\end{align}
while in the T$\sigma$G with  exponential parametrization and C scheme of regularization :
\begin{equation}
	\eta_N^{\text{exp,T$\sigma$G}}=2^{2-d}\pi^{1-\frac{d}{2}}\frac{\Gamma\left(-\frac{d}{2}+m+1\right)}{3\Gamma(m)}\left(d^2-3d-36\right)m^{\frac{d}{2}-1} g_\Lambda \; ,
    \label{nice eta}
\end{equation}
\begin{align}
	\Phi^{\text{exp,T$\sigma$G}}=&2^{2-d} \pi ^{1-\frac{d}{2}} \frac{\Gamma \left(m-\frac{d}{2}\right)}{\Gamma (m)}m^{\frac{d}{2}}d(d-3) g_\Lambda \; .
    \label{nice phi}
\end{align}

Finally, in the B scheme of regularization
$\eta_N$ and $\Phi$ can be expressed in terms of their
counterparts in the C scheme for every gauge and parametrization:
\begin{equation}\label{etaBscheme}
	\eta^{\text{B}}_N=\dfrac{\eta^{\text{C}}_N}{1+\dfrac{\eta^{\text{C}}_N}{2}} \; ,
\end{equation}
\begin{equation}\label{PhiBscheme}
	\Phi^{\text{B}}=\left(1-\dfrac{\eta^{\text{B}}_N}{2}\right)\Phi^{\text{C}} \; .
\end{equation}
\end{appendices}
\bibliography{biblio}
\end{document}